\documentclass[aps,preprint,12pt]{revtex4-2}
\usepackage{amsmath}
\usepackage{amssymb}
\usepackage{color}
\usepackage{soul}
\sethlcolor{yellow}
\usepackage[most]{tcolorbox}
\usepackage{lipsum}

\begin{document}

\title{Probing Intrinsic Elastic Properties of Multilayer Graphene -- a New Mechanical Constant}
\author{Yun Hwangbo}
\author{Seong-jae Jeon}
\affiliation{Korea Institute of Machinery $\&$ Materials, Daejeon 34103, Republic of Korea}
\author{Young-Woo Son}
\affiliation{Korea Institute for Advanced Study, Seoul 02455, Republice of Korea}
\author{Sungjong Woo}
\affiliation{Department of Physics, Pukyong National University, Pusan 48513, Republic of Korea}
\affiliation{Center for Theoretical Physics of Complex Systems, Institute for Basic Science, Daejeon 34126, Republic of Korea}
\begin{abstract}
We present measurements on in-plane Young's modulus and the Gr\"{u}neisen parameter of multilayer graphene with varying number of layers, obtained through {\it in situ} bulge tests.
Accurate determination of their elastic parameters poses a significant experimental challenge due to the substantial differences in mechanical behavior between intra- and inter-layers. To address this, we develop a novel theoretical model with first-principles calculations to investigate thickness-dependent incomplete strain transfer between the layers.
Our findings show that the experimentally measured elastic constants, which deviate from computed intrinsic values, fail to fully capture ideal mechanical couplings between layers. 
As a solution, we propose a new mechanical modulus that integrates the Gr\"{u}neisen parameter and in-plane Young's modulus, providing a more reliable representation of their mechanical properties, independent of unavoidable interlayer effects.
\end{abstract}
\maketitle

\section{Introduction}
Multilayered two-dimensional (2D) materials have highly anisotropic elastic behavior due to their unique atomic structure ~\citep{Novoselov2016Science, Manzeli2017NRM, Wang2025Nanoscale}.
Single layer is formed by covalent bonds while neighboring layers are bound by van der Waals (vdW) interaction. 
The vdW interaction, though is weaker in materials compared with the ionic, covalent and metallic bonds~\cite{Callister2009materials}, is essential for constructing various multilayered 2D materials ~\cite{Geim2013Nature, Gomez2022NatRev}.
In graphite, the strong $sp^2$ covalent bond makes a hexagonal lattice while layers glue together by the weak vdW attraction~\cite{Wang2017PRL}. 
Owing to anisotropic natures in their structures, many physical properties of layered materials such as electrical, thermal and optical ones change their characteristics from their 2D limit to three-dimensional (3D) bulk ones depending on the number of layers ($N_\text{lay}$)~\cite{Dutta2024NanoMatSci, Zhang2025NatNano, Yu2023PRB, Pandey2022optic}.
Their mechanical properties may also alter accordingly. Based on currently available studies~\cite{Lee2009Solidi, WenXing2004Physica, Zhang2012DRM, ZHANG2013CMS,  Lee2012NanoLett, Annamalai_2012, Mortazavi2012CMS, Xiang2015JPD, Han2016EPL, Wang2019PRL, Zhong2019JAP},
the dimensional dependence of mechanical properties of graphene, especially in-plane Young’s modulus ($E$), is not thoroughly investigated.
In-plane Young's modulus is defined by $E=E_\text{2D}t^{-1}$ where $E_\text{2D}$ and $t$ are 2D Young's modulus and thickness of the sample, respectively.
When the thickness of a multilayer graphene increases, all three possibilities for variation of $E$ – increase, decrease or remain unchanged – have been reported.
Questions such as whether there is any dimensional crossover or transition for mechanical properties, and if there is, what is the critical $N_\text{lay}$ and which physical parameter dominates, are important and fundamental but have not been answered conclusively so far.

\begin{figure}
    \centering
    \includegraphics[width=\linewidth]{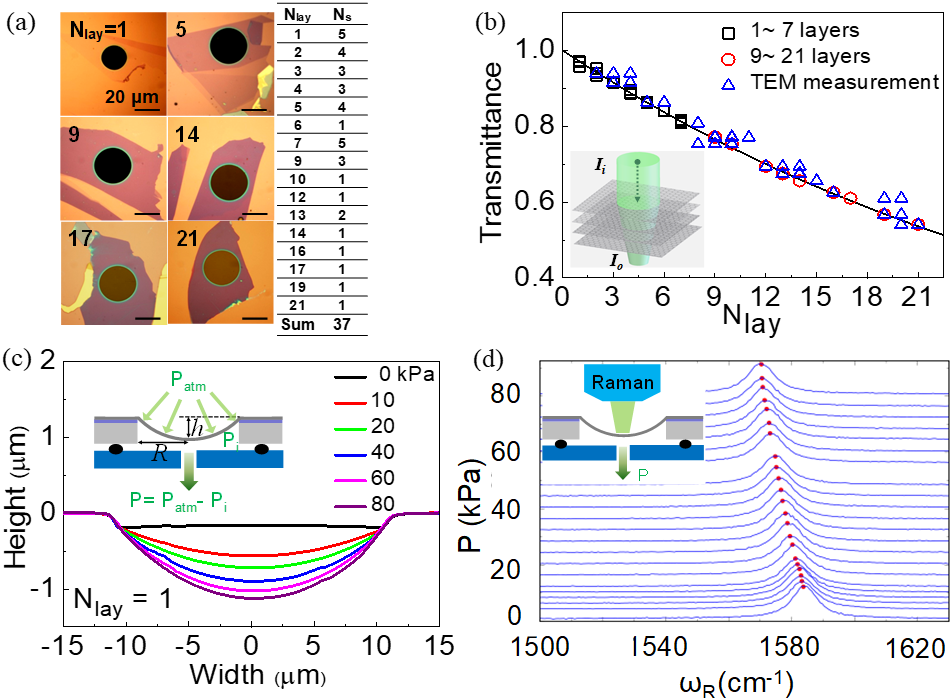}
    \caption{(a) Graphene with various numbers of layers ($N_\text{lay}$) on perforated SiO\textsubscript{2}/Si substrate  with different numbers of samples ($N_S$). (b) Optical transmittance as a function of $N_\text{lay}$. For $N_\text{layer}\le 7$, Raman spectroscopy is used to measure $N_\text{lay}$ and the transmittance for $N_\text{lay}> 7$. (c) {\it In situ} AFM-bulge test data and schematics (inset) and (d) Raman spectra around G peak (red dots) with varying pressures (P).}
    \label{fig:Fig1}
\end{figure}

In this study, we explore the elastic behaviors of graphene such as the dependence of $E$ and Grüneisen parameter ($\gamma$) on the sample thickness using {\it in situ} bulge tests combined with simultaneous spectroscopic measurements. We also perform extensive first-principles calculations based on density functional theory (DFT) and devise a new mechanical model to consider realistic situations beyond simple periodic unitcell geometries of DFT calculations. 
Our main finding is a sharp dichotomy between measurements and DFT results such that 
theoretical $E$~\cite{Woo2016PRB} and $\gamma$ based on DFT are independent of $N_\text{lay}$ while the measured values for both of them substantially decrease with increasing thickness.
Using a newly developed model, we find that the origin of the measured discrepancy is the experimental overestimation of bulge heights and strains due to inevitable interlayer sliding.
To characterize the intrinsic mechanical properties of multilayered graphene, we propose a new mechanical constant that is robust against unavoidable interlayer sliding and thus measurable directly through experiments. We believe that our discovery of the mechanical properties of multilayer graphene would also be useful for other layered vdW materials.

\begin{figure}
\includegraphics[width=\linewidth]{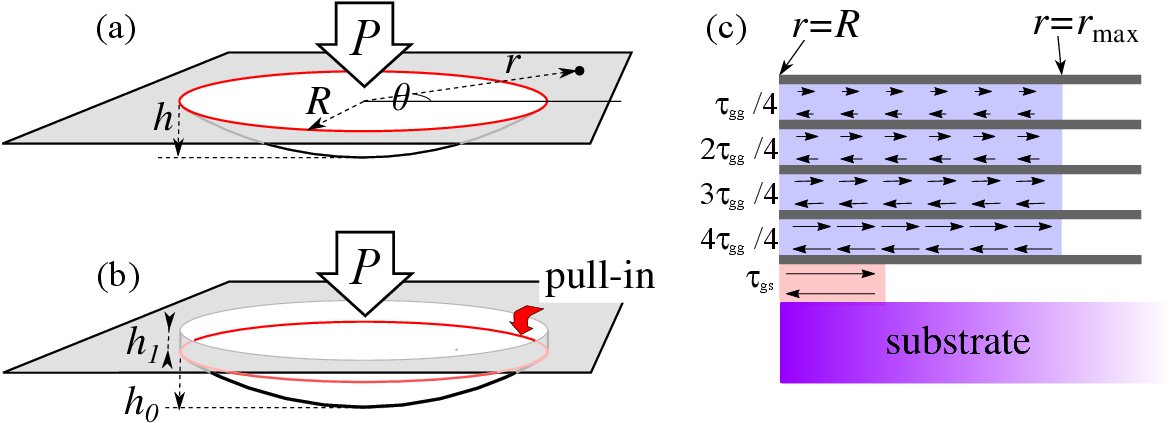}
\caption{(a) Schematic diagram for bulge test with perfect clamping condition or without any sliding. 
$R$ is the radius of bulge hole and $h$ is the bulge height. 
(b) Similar diagram with interlayer sliding. $h_0$ is the same as $h$ in (a) while $h_1$ is the additional height pulled in from the supported region.
(c) Cross-sectional view of the distribution of interlayer stresses in the supported region under an applied pressure of $P$ in case of $N_\text{lay}=5$.
The pairwise arrows pointing opposite directions shown in-between neighboring layers represent interlayer shear stress resisting external pressure. 
For each layer, the force arising from the difference in shear stresses above and below act in positive radial direction and is balanced by the gradient of intralayer stress upto $r=r_{\rm max}$.}
\label{fig:Fig2}
\end{figure}

\section{Materials and experiments}
We prepare graphene samples with various thickness ranging from single to twenty one layers (see section I of SI and Fig.~S1 for details). Figure~\ref{fig:Fig1}(a) shows optical images of representative graphene samples.
To determine $N_\text{lay}$ precisely, we mainly use optical transmittance~\cite{Hwangbo2014Carbon} and Raman spectroscopy shown in Fig.~\ref{fig:Fig1}(b) and (d), respectively ~\cite{Ferrari2006PRL, Li2023Progress}.
Up to seven layers, Raman spectroscopy precisely determines $N_\text{lay}$ and optical  transmittance data are fitted to $N_\text{lay}(>7)$ from the Raman measurement and extrapolated beyond it (see section II, III of SI and Fig.~S2 for further details). TEM~\cite{Ping2012NL} measurements are also carried out to cross-check them (see section II of SI and Fig.~S2C,~S3 for details).

To measure $E$, {\it in situ} bulge tests are performed along with atomic force microscopy (AFM) as shown in Fig.~\ref{fig:Fig1}(c). Stress is applied on graphene using pressure difference ($P$) between the atmosphere ($P_{\rm atm}$) and the vacuum chamber ($P_{\rm i}$) so that $P=P_{\rm atm} - P_{\rm i}$ where $P_{\rm i} < P_{\rm atm}$. Bulge height ($h$) is measured by a non-contact mode of AFM (see section IV of SI and Fig.~S4 for details). 
To obtain $\gamma$, Raman shift is measured from the {\it in situ} Raman-bulge test as shown in Fig.~\ref{fig:Fig1}(d). The positions of Raman $G$ and $2D$ bands are obtained using Lorentzian fitting for each pressure.  Here, only $\gamma$ for the $G$ band is analyzed (see section V of SI and Fig. S5 for details).

From the measured $h$ for a given $P$, the values of biaxial strain ($\epsilon\equiv\epsilon_{xx}=\epsilon_{yy}$) and stress ($\sigma\equiv\sigma_{xx}=\sigma_{yy}$) at the center of the graphene are obtained using $\epsilon = \frac{2h^2}{3R^2}$ and $\sigma = \frac{PR^2}{4ht}$ and then experimental in-plane Young's modulus is typically determined using~\cite{Small1992JMS, Cao2023JPhysD} 
\begin{align}
\label{simpleFormE}
E = (1-\nu){\frac{\sigma}{\epsilon}} = (1-\nu)\left(\frac{3R^4}{8t}\right)\left(\frac{P}{h^3}\right),
\end{align} 
where $R$, $t$ and $\nu$ are the radius of the bulge hole, the thickness and the Poisson's ratio of the sample as in Fig.~\ref{fig:Fig2}(a).
The expressions for strain ($\epsilon$) and stress ($\sigma$) are derived under the assumption that the bulge adopts a spherical geometry and that $R \gg h, t$. This approximation breaks down near the periphery of the bulge, where strain and stress become highly anisotropic.
Equation~\eqref{simpleFormE} can also be rewritten as
\begin{align}
\label{Eq_hvsq}
P = \left(\frac{8t}{3R^4}\frac{E}{1-\nu}\right) h^3 \equiv kh^3.
\end{align}

\begin{figure}
  \includegraphics[width=\linewidth]{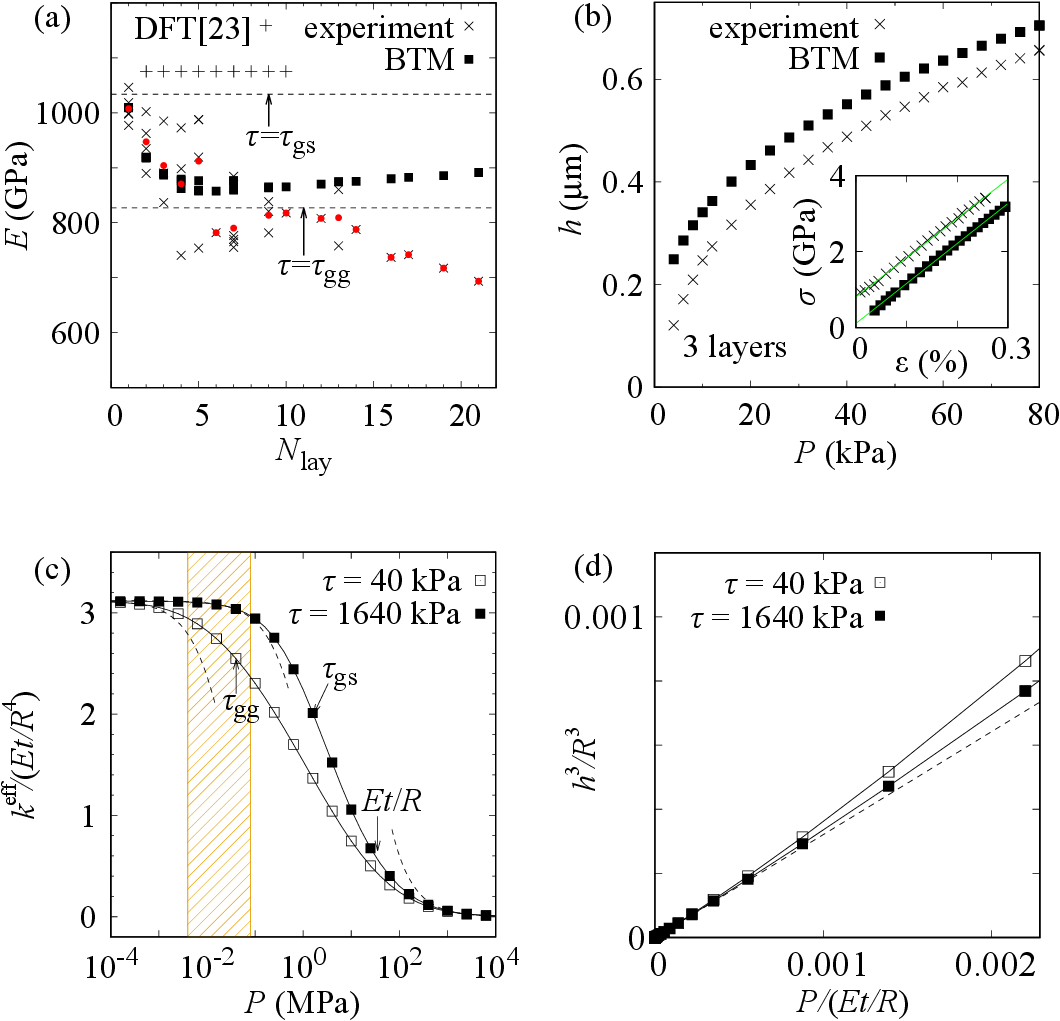}
  \caption{
  (a) Values of $E$ from experiments, DFT calculations, and the BTM model as a function of $N_\text{lay}$. Dashed lines are the theoretical boundaries for $E_\text{ap}$ of a single layer on SiO$_2$ and on graphene. Red solid circles represent the average of measured values for each $N_{\rm lay}$.
  (b) $h$ as a function of $P$ for a trilayer graphene. Inset represents $\sigma$ vs. $\epsilon$ curve to obtain Young's modulus, $E$.
  (c) $k^\text{eff}$ as a function of $P$ for a single-layer graphene with $\tau = \tau_{\rm gg}$ 
  and $\tau_{\rm gs}$. The hatched area is our experimental pressure range.
  (d) Nonlinear behavior of $h^3$ with respect to $P$ depending on $\tau$ for a single-layer graphene. Dashed line is for a perfect clamping case.
  }
  \label{fig:Fig3}
\end{figure}

The diameter of the holes for our samples ranges from 20 to 30 $\mu$m.
Using Eq.~\eqref{simpleFormE}, measured $E$'s depending on $N_{\rm lay}$ from our experiments are plotted in Fig.~\ref{fig:Fig3}(a).
It shows that $E$ decreases substantially with increasing $N_{\rm lay}$.
On the other hand, a previous DFT report shows that the in-plane Young's modulus of multilayer graphene --  also plotted in Fig.~\ref{fig:Fig3}(a) -- is barely dependent on $N_{\rm lay}$~\cite{Woo2016PRB}. 
We will show that the experimental deviation from the DFT result can be attributed to the imperfect clamping of the graphene to the substrate as well as the interlayer sliding between graphene layers.

With interlayer sliding considered, interlayer shear stress ($\tau$) is built against the direction of sliding, that can be distinguished for one ($\tau_{\rm gs}$) between graphene and substrate, and the other ($\tau_{\rm gg}$) between adjacent graphene layers. 
When the suspended graphene within a hole in the substrate is subjected to pressure, a constant shear stress develops due to interlayer sliding in the supported region. This shear stress equilibrates with the gradient of intralayer stress, which is further balanced against the intralayer stress of the suspended graphene.
It is important to note that in a bulge test, the shear stress exceeds the linear regime and is limited by a maximum value specific to the material properties.
A previous success~\cite{Wang2017PRL} for $N_\text{lay}\le 2$ using a constant $\tau$ may be limited to thin samples since the $\tau$ should be distributed properly across the layers for thicker ones, as will be discussed below. 
To address cases with more than two layers, we propose a new mechanical model for Bulge Test of Multilayered vdW materials (we will call this the BTM model hereafter). This model estimates the extent to which the sample near the rim of the hole is pulled inward under single and multilayer conditions. Such inward pulling adds to the bulge height, making the sample appear softer.

\section{Theoretical model}
To describe the BTM model, let us begin with the simplest case, a single-layer graphene.
The equation of equilibrium of graphene in the supported region ($r>R$) is
\begin{align}
\label{eqn of equil rt}
\frac{\partial\sigma_r}{\partial r} + \frac{\sigma_r -\sigma_\theta}{r} + 
\frac{\tau}{t} = 0.
\end{align}
The stress for the suspended region ($r<R$) is obtained by $\sigma = \frac{PR^2}{4ht}$.  
Near $r\sim R$ in the suspended region, however, $\sigma_\theta$ becomes negligible
so that $\sigma_r$ should become roughly doubled in order to bear the same pressure.
The values of $\sigma_r$ and $\sigma_\theta$ in the supported region can be obtained using Eq.~\eqref{eqn of equil rt} with the continuity of $\sigma_r$ at $r=R$ as well as another boundary condition of simultaneous vanishings of $\sigma_r$ and $\sigma_\theta$ at $r=r_\text{max}(>R)$.
It should be noted that $\sigma_\theta$ need not be continuous across $r=R$.
It turns out that these are determined as a function of a dimensionless parameter of $\zeta\equiv\left[\frac{\left(P/\tau\right)^2(E/\tau)}{R/t}\right]^{1/3}$ (see E of section VI in SI for details).
The azimuthal strain can be obtained accordingly using $\epsilon_\theta = \left(\sigma_\theta - \nu\sigma_r \right) /E$.
With the isotropy condition, the radial displacement in the supported region can also be obtained as $u_r = r\epsilon_{\theta}$.
The $u_r$ at $r=R$ represents the {\it amount of pulling-in} of graphene at the rim of the hole.
As depicted schematically in Fig.~\ref{fig:Fig2}(b), this pulling-in provides additional contribution to the bulge height, $h_1 = -u_r(r=R)$, so that the total bulge height becomes $h = h_0 + h_1$, where $h_0$ is from Eq.~\eqref{Eq_hvsq}.
(see D, E of section VI in SI for futher details).

Now, let us extend our BTM model to multilayer cases, in which shear stress must be distributed properly across the layers with the maximal value limited by $\tau_{gg}$ or $\tau_{gs}$.
Figure~\ref{fig:Fig2}(c) shows the configuration of the distribution of interlayer shear stresses; the difference between neighboring $\frac{\tau}{t}$'s should be 
equilibrated with the gradient of intralayer stress of each layer in the supported region.
For static equilibrium, the values of $h$ for all the layers should be the same.
Such an equilibrium condition can be achieved if the intralayer strain (or stress) distributions are the same for all the layers in the supported region.
It allows $r_{\rm max}$ of each layer and corresponding $h$ to be all the same in case that the total pressure is equally distributed across the layers (see G of section VI in SI for further discussion about the multilayer equilibrium condition as well as the role of $\tau$).
The same intralayer stress distribution for all the layers demands that the differences between neighboring $\tau$'s are to be the same; $\tau$ must increase linearly from the top to lower layers.
The bottom layer is an exception since $\tau_{\rm gs}\ne \tau_{\rm gg}$.
Such an equilibrium condition is one of the main advances from the approach of {\it Wang~et.al}~\cite{Wang2017PRL}.
For the bottom layer, $(\tau_{\rm gs}-\tau_{\rm gg})/t$ should be balanced against the radial gradient of intralayer stress.
The total pressure of $P$ is distributed such that $P = (N_{\rm lay}-1)P_{\rm u} + P_{\rm b}$. 
Here, the ratio between $P_{\rm u}$ and $P_{\rm b}$ is determined to make the values of $h$ for all the layers be the same.
The bulge height for the $i$-th layer for a given pressure of $P_i$ is calculated using the same procedure as in the single layer case and the subscripts `u' and `b' represent the upper and bottom layers, respectively.
Detailed derivation of the BTM model is provided in section VI of SI.

\section{Results and discussion}
\subsection{Measured \& apparent Young's modulus}
We calculate $h$ based on the BTM model using theoretical $E$ from DFT~\cite{Woo2016PRB}.
Figure~\ref{fig:Fig3}(b) compares values of $h$ based on the BTM model with those from experiments as a function of $P$ for trilayer graphene.
The inset shows corresponding $\sigma$ vs. $\epsilon$ plots obtained using the simple formula in  Eq.~\eqref{simpleFormE}.
We note that a constant offset between the two data points does not affect the slope of $\sigma$ vs. $\epsilon$ nor the computed in-plane Young's modulus of $E$.
Let us call this Young's modulus obtained with a simple ratio of $(1-\nu){\frac{d\sigma}{d\epsilon}}$ from our BTM model as `apparent Young's modulus' or $E_\text{ap}$ as compared to the intrinsic $E$. 
The solid squares in Fig.~\ref{fig:Fig3}(a) are the $E_\text{ap}$ using the sample geometries, while the $\times$ marks are directly from the experiments.
One can see that $E_\text{ap}$ decreases as $N_\text{lay}$ increases and saturates around $N_\text{lay}\simeq 10$.
It matches pretty well with averaged experimental results (solid red circles).{\footnote{The values of experimental data are quite scattered for each $N_{\rm lay}$ which is attributed to the complex experimental conditions that our mechanical model cannot fully capture. Few of the reasons are as follows. The narrow experimental range of applied stress due to the limit of maximum pressure is one of the factors -- the pressure in our experiments is upper limited by the atmospheric pressure. Our choice of the hole size of tens of micrometers, which is the marginal size for reliable experiments, is for pushing up the pressure range. The measurement of the initial zero-point height of the bulge also provides some level of uncertainty since the shape of bulge is not perfectly flat or spherical. It is the reason why we use many samples so that the uncertainty would be systematically reduced. In consequence, the average value of Young's modulus for each $N_{\rm lay}$ does follow the mechanical model prediction reasonably well.}
However, for samples thicker than ten layers, they deviate significantly.
We consider ten layers as a critical thickness for mechanical crossover from 2D to 3D characteristics as will be discussed further later in this paper.

\subsection{Saturation of apparent Young's modulus}
The saturation of $E_{\rm ap}$ with increasing $N_{\rm lay}$ comes from the fact that there are two values for interlayer shear stress, $\tau_{gs}$ and $\tau_{gg}$. 
In order to understand this, let us reconsider the single layer case with Eq.~\eqref{Eq_hvsq}.
With modified $h=h_0+h_1$, the ratio $k = \frac{P}{h^3}$ in Eq.~\eqref{Eq_hvsq} can be shown to be no more constant but to depend on a dimensionless parameter,
$\xi\equiv\left[\frac{(R/t)^2}{(P/
  \tau)(E/\tau)^2}\right]^{1/3}$ such as
\begin{eqnarray}
    k^\text{eff}=\frac{Et}{R^4}\left[
    \frac{\left(3(1-\nu)\right)^\frac{1}{3}}{2} 
    + \xi\frac{(1-\nu^2)}{6} f\left(\frac{r_\text{max}}{R}\right)
    \right]^{-{3}},
    \label{Eq_kzeta}
\end{eqnarray}
where $f(x)\equiv 
\left(x-1\right)^3+ 3\left(x-1\right)^2$ and $r_\text{max}$ depends on $\zeta$ (see SI for details).
Noting that $\zeta/\xi=\frac{(P/\tau)(E/\tau)}{R/t}$, $k^\text{eff}$ can be approximated for two limiting cases of $\zeta/\xi\ll 1$(low $P$) and $\zeta/\xi \gg 1$ (high $P$) with $\nu=0.144$ ~\cite{Woo2016PRB} as
\begin{align}
\frac{k^\text{eff}}{Et/R^4} 
\label{Eq_criteria1}
\approx 3.12
\left(1 + \frac{P}{\tau}\frac{(1+\nu)}{3}
\right)^{-3},
\end{align}
and
\begin{align}
\label{Eq_criteria2}
\frac{k^\text{eff}}{Et/R^4} 
\approx 1.72\left(\frac{Et}{R}\right)\frac{1}{P},
\end{align}
respectively.
We note that $k^\text{eff}$ begins to decrease from the perfect clamping case if $P \sim \tau$ and approaches to zero if $P \gg Et/R$.
These two criteria determine the overall dependence of $k^\text{eff}$ on $P$.
From our experimental geometry of $R\sim 10~\mu$m and $t = 3.35$~{\AA}, the high pressure criterion is estimated to be $P \gg 35$~MPa well above our applied pressure of orders of 10 kPa.
So, the low pressure limit fits our cases (see F of section VI in SI for detailed derivations).
Figure~\ref{fig:Fig3}(c) shows $k^\text{eff}$ as a function of $P$ for $\tau_{\rm gg} = 40$~kPa and $\tau_{\rm gs} = 1640$~kPa~\cite{Wang2017PRL}. 
It begins to decrease at different pressures corresponding to each $\tau$ value, while it supresses to zero over $P\sim Et/R$.
The hatched area is the experimental pressure range.
Figure~\ref{fig:Fig3}(d) shows the deviation from the linear correlation (dotted line) between $h^3$  and $P$  upto the hatched area in Fig.~\ref{fig:Fig3}(c).
The maximal amounts of decrease in $k^\text{eff}$ within the experimental pressure range are 4~\% and 22~\% for $\tau_{gs}$ and $\tau_{gg}$, respectively.
Since $E_\text{ap}$ is proportional to $k^\text{eff}$ in Eq.~\eqref{Eq_hvsq}, {\it i.e.}, $E_\text{ap}=\frac{3R^4}{8t}({1-\nu})k^\text{eff}$, it also decreases to $1034$ and $827$~GPa for $\tau_{gs}$ and $\tau_{gg}$, respectively, from its intrinsic value of $1075$~GPa.
For a multilayer sample, $E_\text{ap}$ for each layer should be averaged so that, as $N_{\rm lay}$ increases, its overall value approaches one for $\tau_{\rm gg}$ from another for $\tau_{\rm gs}$.
In Fig.~\ref{fig:Fig3}(a), $E_\text{ap}$ is shown to start with the value for $\tau_{\rm gs}$ and approaches that for $\tau_{\rm gg}$ as $N_{\rm lay}$ increases.

\subsection{2D to 3D crossover}
For the deviation of $E_\text{ap}$ from the experimental results for $N_\text{lay}>10$,
we suspect two potential interrelated sources, wrinkle formation~\cite{Han2016EPL, Zhong2019JAP, Ares2021PNAS} and sample delamination. 
Because $u_r = r\epsilon_{\theta}$, isotropic pull of graphene toward the center of the hole causes azimuthal compression~\cite{Ares2021PNAS}, forming wrinkles or local delamination especially near the edge of the hole where the pressurized graphene is bent and structurally unstable.
Such instability is expected to increase with thickness.
Pressure-induced wrinkles, which are aligned radially, effectively reduce the local elastic modulus of $E$ along the circumferential direction. This alteration shifts one of the pressure criteria, $Et/R$, toward lower pressure of $P$, thereby changing the shape shown in Fig.~\ref{fig:Fig3}(c).
Furthermore, local structural failure due to the wrinkles might result in delamination especially for thicker samples, causing reduction of $\tau_{gs}$.
These two effects can further reduce $k^\text{eff}$ as well as $E_{\rm ap}$ beyond our BTM model even though a rigorous quantitative analysis is prohibitive within our modeling approach due to the complication of the experimental systems.
A material could be classified as two-dimensional if it exhibits a relatively large transverse degree of freedom. In the case of multilayered structures, the BTM model effectively captures their behavior within the 2D regime by accounting for transverse deformations coupled with interlayer sliding. We propose that the breakdown of the BTM model for thicker multilayer systems signifies a dimensional crossover from two- to three-dimensional behavior, which occurs around a critical number of layers, approximately $N_\text{lay} \simeq 10$.

\begin{figure}
    \includegraphics[width=\linewidth]{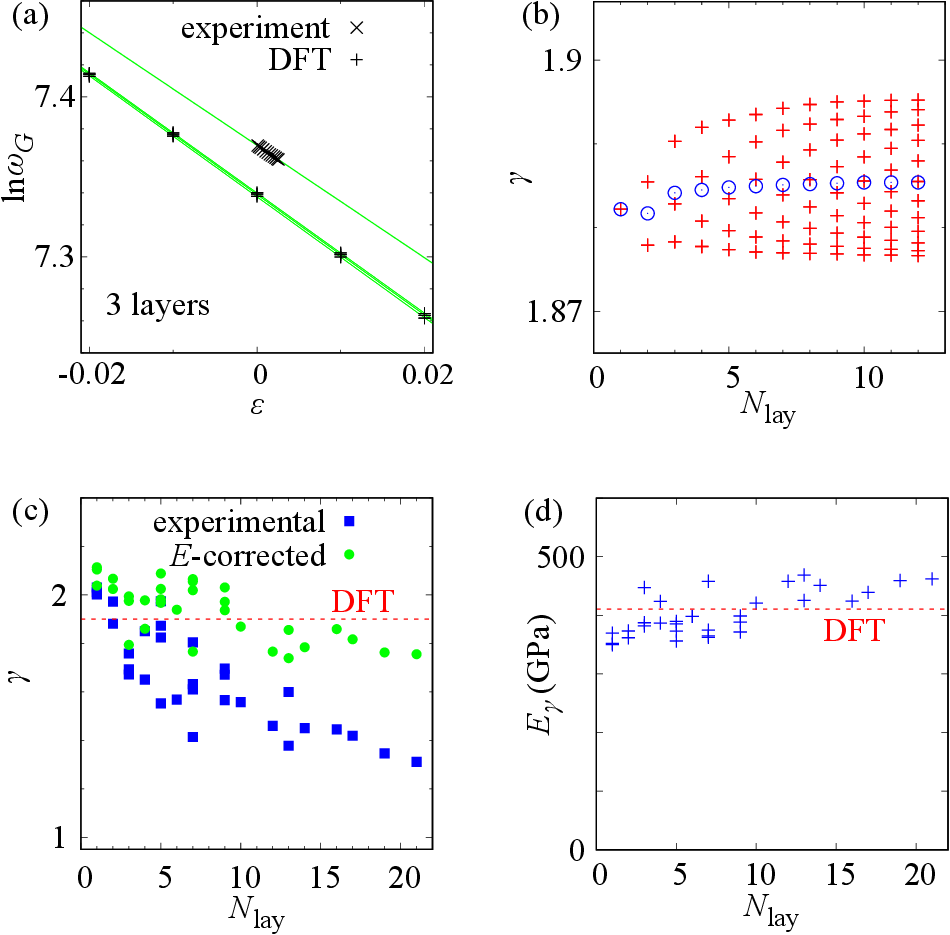}
    \caption{(a)
    Computed $\ln\omega_G$ as a function of  $\epsilon$ from DFT calculation compared with measured Raman $G$ peak for graphene for $N_\text{lay}=3$. There are six modes from DFT for each $\epsilon$. Green lines show linear fitting curves.
    (b) Theoretical $\gamma$ from DFT calculations for varying $N_\text{lay}$.
    (c) Measured Grüneisen parameter of $\gamma$ for varying $N_\text{lay}$ (blue solid rectangles). Green solid circles represent $\tilde{\gamma}$ corrected from measured $\gamma$.
    (d) Measured elastic Gr\"{u}neisen modulus ($E_\gamma$) with varying $N_\text{lay}$ (blue crosses) and the constant red dashed line from DFT.}
    \label{fig:Fig4}
\end{figure}

\subsection{Measured Gr\"{u}neisen parameter}
To further investigate the overestimation in bulge height and strain induced by the effect of wrinkles and delamination as well as the interlayer sliding, we have obtained Grüneisen parameter of $\gamma$ using Raman spectrum data.
We have also obtained theoretical $\gamma$ based on DFT calculations~\cite{Giannozzi_2009} for comparison.
In order to properly consider van der Waals interaction between layers, we have used the revised version~\cite{Stefano2013PRB} of the nonlocal correlation functional as in ref~\cite{Woo2016PRB} developed by Vydrov and van Voorhis~\cite{Troy2010JChemPhys} which is known to be good for predicting the interlayer distance and phonon frequencies of graphite~\cite{Stefano2016PRB}.
Phonon frequencies of $\omega_G$ corresponding to the Raman $G$ peak are calculated for multilayer graphene samples under biaxial strain varying from $-0.02$ to $0.02$.
Measured and calculated $\ln\omega_G$ with varying $\epsilon$ are drawn in Fig.~\ref{fig:Fig4}(a) for $N_\text{lay}=3$ (see section VII of SI for details).
Using $\gamma=-(1/2){\partial\ln\omega_G}/{\partial\epsilon}$~\cite{Mounet2005PRB,Huang2016CMS}, we obtain $\gamma$ for doubly degenerate $2N_{\rm lay}$ modes corresponding to the Raman $G$ peak~\cite{Mohiuddin2009PRB, Cheng2011PRB, Androulidakis2015SciRep}~[Fig.~\ref{fig:Fig4}(b)].
We note that the splittings are less than one percent of the average values while the shift of average values depending on $N_{\rm lay}$ is at most a few tenth percent, so that within experimental resolution $\gamma$ can be considered almost constant.
For $N_\text{lay}=1$, our DFT result agrees well with a previous report~\cite{Androulidakis2015SciRep}.
In Fig.~\ref{fig:Fig4}(c), the measured $\gamma$ (blue squares) from Raman spectrum are shown to decrease as $N_\text{lay}$ increases in contrast to the constancy of averaged $\gamma$ from the DFT calculation.
We find that this discrepancy is also attributed to the overestimation of bulge height and strain as explained below.

\subsection{Elastic Gr\"{u}neisen modulus}
Let $\tilde{\epsilon}$ and $\tilde{\sigma}$ be the intrinsic strain and stress averaged over the layers.
The intrinsic strain $\tilde{\epsilon}$ is determined by $h_0$ instead of $h$ such that $\tilde{\epsilon} = (2/3)(h_0/R)^2$. 
However, $h_0$ cannot be extracted directly from the experiment.
Instead, since the intrinsic in-plane Young's modulus of $\tilde{E} = (1-\nu)\frac{3R^4}{8t}\frac{P}{h_0^3}=1075$~GPa is independent of $N_\text{lay}$~\cite{Woo2016PRB}, $\tilde{\epsilon}$ can be estimated from measured $\epsilon$ using measured $E_{\rm ap} = (1-\nu)\frac{3R^4}{8t}\frac{P}{h^3}=\left(\frac{h_0}{h}\right)^3\tilde{E}$ such that
\begin{align}
    \tilde{\epsilon} 
    = \frac{2}{3}\left(\frac{h_0}{R}\right)^2 
    = \frac{2}{3}\left(\frac{h}{R}\right)^2\left(\frac{h_0}{h}\right)^2 
    = \epsilon \left(\frac{E_{\rm ap}}{\tilde{E}}\right)^{\frac{2}{3}}.
\label{eq:re_epsilon}
\end{align}
This formula shows a quantitative scale for the overestimated  $\epsilon$ using the measured (hence, underestimated) $E_{\rm ap}$. 
Similarly, the intrinsic Gr\"{u}neisen parameter of $\tilde{\gamma}$ (also independent of $N_\text{lay}$) can be expressed as
\begin{equation}
\tilde{\gamma} =
\frac{\partial\ln\omega}{2\partial\tilde{\epsilon}} = 
\frac{\mathrm{d}\epsilon}{\mathrm{d}\tilde{\epsilon}}
\frac{\partial\ln\omega}{2\partial \epsilon} = \left(\frac{\tilde{E}}{E_{\rm ap}}\right)^\frac{2}{3}\gamma.
\label{eq:re_gamma}
\end{equation}
Here, we note that the measured $\gamma$ is underestimated by the same factor for the overestimated strain in Eq.~\eqref{eq:re_epsilon}.
Equation~\eqref{eq:re_gamma} can be used to correct measured $\gamma$ to intrinsic $\tilde{\gamma}$, by substituting experimental $E_{\rm ap}$.
Figure~\ref{fig:Fig4}(c) shows that $\tilde{\gamma}$ (solid green circles) agrees reasonably well with the DFT data.
The correction also reduces the dispersion of data points for each $N_{\rm lay}$, confirming that the observed variability in experimental $E$ and $\gamma$ is largely due to the common factor of unavoidable overestimation of bulge height and strain. This overestimation is primarily caused by the interlayer sliding, although some aspects of its detailed mechanism lie beyond the BTM model, especially for $N_\text{lay}>10$.

By noting that $\tilde{E}\tilde{\gamma}^{-\frac{3}{2}}=E_\text{ap}\gamma^{-\frac{3}{2}}$ from Eq.~\eqref{eq:re_gamma},
we propose a new mechanical constant, termed the \textit{elastic Gr\"{u}neisen modulus}, defined as \( E_\gamma \equiv E \gamma^{-3/2} \). This quantity can be directly extracted from experimental measurements and represents an intrinsic material property that remains robust against unavoidable strain overestimations encountered in experimental setups.
In Fig.~\ref{fig:Fig4}(d), it is shown that measured $E_\gamma$ values from our experiment are indeed nearly independent of sample thickness and 
match well with the constant DFT result.

\subsection{Generality and significance}
Before conclusion, we would like to briefly mention the generality of the theoretical method and the significance of the proposed constant in terms of various materials. The interlayer sliding effect in measuring mechanical moduli of other vdW materials such as BP, BN, $h$BN, MoS$_2$, WSe$_2$, WTe$_2$ etc., have been conjectured in many experiments, although the detailed mechanism has yet to be clearly revealed~\cite{Zhang2022JPCC, Deng2022npj, Wang2023IntJExtrem, Sun2022Prog, Gon2012ACSnano}.
Our proposition of detailed theoretical mechanism of interlayer interaction is unprecedented as far as we know. Investigation on various materials in terms of the new mechanical constant is therefore important and should be included in the future project.

\section{Conclusion}
We have measured and analyzed important elastic properties of multilayer graphene such as in-plane Young's modulus and Gr\"{u}neisen parameter carefully.
Unlike the thickness-independent values obtained from DFT calculations, our measurements reveal a strong dependence on the number of layers, which we attribute primarily to the unavoidable interlayer sliding.  
Additionally, we introduce a new material constant, the elastic Gr\"{u}neisen modulus, which remains experimentally robust despite interlayer sliding effects. 
Given that similar mechanical properties are observed across nearly all vdW materials
we believe that our new model and constant will play a crucial role in accurately extracting intrinsic mechanical characteristics from experiments. 

\section*{Acknowledgements}
Y.H.  and S.-j.J. were supported by the National Research Foundation (NRF) funded by the Korean government (MSIT) (RS-2024-00432936). Y.-W.S. was supported by KIAS individual Grant (No.CG031509).
S.W. was partially supported by the National Research Foundation of Korea (NRF) grant funded by the Korea government (MSIT) (RS-2023-00278511). Computations were supproted by CAC of KIAS.

\section*{Data availability}
Data will be made available on request.

\bibliography{elastic}
\bibliographystyle{elsarticle-num}
\end{document}